\documentclass[12pt, titlepage]{utarticle}
\usepackage{amssymb, amsmath, amsopn, amsthm,amscd,amsfonts}
\usepackage{graphicx}

\numberwithin{equation}{section}

\newcommand{\e}{\epsilon}

\begin{document}
\preprint{UTTG-10-09 \\
MPP-2009-34\\
}

\title{Hagedorn Systems from Backreacted Finite Temperature $N_f=2 N_c$ Backgrounds}

\author{Elena C{\'a}ceres,
	 \address{
	  Facultad de Ciencias,\\
      Universidad de Colima, \\
      Bernal D{\'i}az del Castillo 340,\\
      C.P. 28045 Colima, Colima, M{\'e}xico\\
     }
	Raphael Flauger
	 \address{
	  Theory Group, Department of Physics,\\
      University of Texas,\\
      Austin, TX 78712, USA\\
      }
    and Timm Wrase
     \address{
      Max-Planck-Institut f\"ur Physik,\\
      F\"ohringer Ring 6,\\
      80805 M\"unchen, Germany\\
      {~}\\
      {\rm Email}\\
      \emailt{elenac@ucol.mx}\\
      \emailt{flauger@physics.utexas.edu}\\
      \emailt{wrase@mppmu.mpg.de}}
}

\Abstract{ We present non-extremal generalizations of  $N_f=2 N_c $ supergravity duals obtained by adding flavor branes to the Chamseddine-Volkov-Maldacena-Nu\~nez background and taking their backreaction into account. For a particular range of parameters we obtain a one parameter family of analytic black hole solutions with regular horizons. These solutions have a temperature that is always above the Hagedorn temperature of little string theory. We find that including flavor branes and their backreaction increases the minimum temperature of these systems from $T = \frac{1 }{4 \pi}$, for the background without flavor, to $T = \frac{1}{2 \pi \sqrt{1+ \e}}, \, \e\ll1$. We show that outside the range of validity of the analytic solutions presented here the backgrounds have naked singularities. }

\maketitle

\section{Introduction}
The gauge/gravity duality (\cite{Maldacena:1997re},\cite{Gubser:1998bc},\cite{Witten:1998qj}) has proven to be a useful tool to study strongly coupled gauge theories. In the last few years  finite temperature backgrounds have been used to explore strongly coupled plasmas yielding promising results. However, none of these non-extremal backgrounds includes matter in the fundamental representation other than in the quenched limit, for which $\frac{N_f}{N_c}\rightarrow 0$ as $N_c \rightarrow \infty$. Even at $T=0$ the issue of how to account for the  backreaction of a large number of flavors is still being investigated. An important step in that directions was  \cite{Casero:2006pt} where the authors presented a novel setup to
include $N_f$ flavor branes, with $\frac{N_f}{N_c}\sim O(1)$,  in the Chamseddine-Volkov-Maldacena-Nu\~nez (CVMN) model \cite{Maldacena:2000yy}. See also (\cite{Benini:2006hh}-\cite{Gaillard:2009kz})  for work that, using the techniques presented in  \cite{Casero:2006pt}, studies the inclusion of a large number of flavor branes in other backgrounds.\\
The CVMN background represents a large number of NS5 branes wrapped on a two-cycle of the resolved conifold. In the infrared it is dual to ${\mathcal N=1 }\ \  SU(N)$ supersymmetric  Yang-Mills in four dimensions. The ultraviolet completion is  not a field  theory but a little string theory  (\cite{Berkooz:1997cq}, \cite{Seiberg:1997zk}). The non-extremal deformation of CVMN  was studied in \cite{Gubser:2001eg}. It was shown to have  Hagedorn behavior and to be  thermodynamically unstable at high energies, behaving very much like finite temperature LST.\\
In \cite{Caceres:2007mu}\footnote{The reader interested in a summary of the main results is referred to \cite{Caceres:2009zz}.} we found a one parameter family of solutions representing $N_f= 2N_c$ flavor branes and their backreaction in the CVMN background. These solutions are conjectured to be dual to an ${\mathcal N}=1$ $SU(N)$ theory  with $N_f= 2N_c$ matter  in the fundamental representation and some extra matter in the adjoint representation.\\
In the present note we find the non-extremal generalization of the solutions obtained in \cite{Caceres:2007mu}. Our motivation is three-fold: 1)we want to investigate if turning on the temperature hides  the singularity of the solutions  presented in \cite{Caceres:2007mu}, 2) if that is the
case, we want to understand if  these backgrounds are  good duals to finite temperature field theories or if they inherit the thermodynamical behavior of little string theory, 3) if they do behave like finite temperature LST, we want to understand  what is the effect of the flavor branes and their backreaction.\\
For certain range of parameters we  find a one parameter family of analytic solutions with regular horizons. The temperature of these solutions is $T \gtrsim \frac{1}{2 \pi \sqrt{1+ \e}}, \, \e\ll1$  indicating that its temperature is always above the Hagedorn temperature of LST which is $T=\frac{1}{4 \pi}$. Outside  the range of validity of our analytic solutions we have to perform a numerical study and find that the solutions have naked singularities. Hence, we conclude that our black hole solutions are not dual to a finite temperature field theory but to a little string theory with a temperature above the Hagedorn temperature.\\
Only few examples of supergravity backgrounds with Hagedorn behavior are known: NS5 branes and wrapped NS5 branes (\cite{Gubser:2001eg}, \cite{Buchel:2001dg}). The thermal behavior of  these examples has been thoroughly studied and it taught us several things about little string theory. We know that they are thermodynamically unstable and in \cite{Kutasov:2000jp} it was argued that the instability is of stringy origin. A different explanation was advocated in \cite{Rangamani:2001ir}  where the authors  argued that the instability was of supergravity origin, a Gregory-Lafflamme like instability. This issue, for the case of wrapped NS5 branes, was elucidated in \cite{Buchel:2001dg} where it was shown that the instability is indeed of supergravity origin. We  expect that our solutions will also be thermodynamically unstable, but  further research is required to confirm this.\footnote{ Recently, the thermodynamics of a linear dilaton black hole  was studied in \cite{Bertoldi:2009yi}. Unlike the case studied in \cite{Bertoldi:2009yi},  the solutions presented here are non-extremal deformations of  backgrounds where the field theory degrees of freedom are decoupled,{\it i.e.} are candidates to be gravity duals to finite temperature field theories  in four dimensions.  It would be interesting to understand if the marginal stability found in  \cite{Bertoldi:2009yi} also applies to the solutions presented here.}

\section{Black hole solutions}
In this section we will review the metric ansatz considered in \cite{Gubser:2001eg} which contains finite temperature solutions of the Chamseddine-Volkov-Maldacena-Nu\~nez (CVMN) background, \cite{Maldacena:2000yy}. It turns out that the metric ansatz is general enough to contain finite temperature generalizations of the more general supergravity backgrounds discussed in \cite{Casero:2006pt}, \cite{Casero:2007jj}, \cite{HoyosBadajoz:2008fw} and \cite{Caceres:2007mu}. These more general supergravity backgrounds were found by adding a large number of smeared 'flavor' D5 branes ($N_f\sim N_c$) to the CVMN background and taking their backreaction into account. They are conjectured to be dual to an $\mathcal{N}=1$ $SU(N_c)$ gauge theory with $N_f$ quarks in the fundamental representation.\\
In the next subsections we derive the equations of motion for the fields for the special case of $\frac{N_f}{N_c}=2$ discussed in \cite{Caceres:2007mu}. We find analytic solutions for black holes that have a horizon at large values of the radial coordinate and asymptote to the decoupled one parameter family of solutions discussed in \cite{Caceres:2007mu}. Then we calculate the temperature for these solutions and argue that they are the only black holes within the validity of the supergravity approximation.

\subsection{The metric ansatz and the equations of motion}
The metric ansatz considered in \cite{Gubser:2001eg} reads in Einstein frame\footnote{Note that we have switched the signs of $Y_7$ and $Y_8$ as compared to \cite{Gubser:2001eg}.}
\begin{align}\label{eq:metric}
ds^2 &=  -Y_1\,dt^2+Y_2\,d{\rm x}^n d{\rm x}^n + Y_3\, d\rho^2 + Y_4\,(d\theta^2 +\sin{\theta}^2 d\varphi^2)\nonumber\\
&+ Y_5\, \left( (w^1 + Y_7 d\theta)^2 +(w^2 - Y_7 \sin \theta d \varphi)^2
\right) + Y_6\,(w^3 + \cos \theta d \varphi)^2 \, ,
\end{align}
where $Y_i=Y_i(\rho)$ and $w^a$ are the left-invariant one-forms on the $S^3$
\begin{align}
w^1 &=\cos \psi d\tilde{\theta} + \sin \psi \sin \tilde{\theta} d \tilde{\varphi},\nonumber\\
w^2 &=-\sin \psi d\tilde{\theta} + \cos \psi \sin \tilde{\theta} d \tilde{\varphi},\\\nonumber
w^3 &=d\psi + \cos \tilde{\theta} d \tilde{\varphi}.
\end{align}
In addition we make the following ansatz for the RR 3-form $F_3$ \cite{Casero:2006pt}
\begin{align}
F_{(3)}& = \frac{N_c}{4} \Bigg[ -(w^1 + Y_8 d\theta)\wedge (w^2 - Y_8 \sin \theta d\varphi) \wedge (w^3 +\cos \theta d \varphi) \nonumber\\
& + Y_8' dr \wedge (-d \theta \wedge w^1 + \sin \theta d\varphi \wedge w^2) + \left( 1-Y_8^2-\frac{N_f}{N_c} \right) \sin \theta d \theta \wedge d\varphi \wedge w^3\Bigg].
\end{align}
Note that $F_3$ is not closed due to the presence of the smeared flavor branes. Furthermore, we assume that the dilaton is also a function of $\rho$ and define $\phi = Y_9(\rho)$.\\
The complete action for type IIB supergravity with smeared flavor branes reads
\begin{equation}
S=S_{\text{grav}}+ S_{\text{flavor}},
\end{equation}
where, in Einstein frame, we have
\begin{equation}
S_{\text{grav}}= \frac{1}{2\kappa_{(10)}^2}\int d^{10}x \sqrt{-g_{(10)}} \left( R -\frac{1}{2} (\partial_{\mu} Y_9) (\partial^{\mu} Y_9) - \frac{1}{12}e^{Y_9}F_{(3)}^2 \right),
\end{equation}
and
\begin{equation}
S_{\text{flavor}} =\frac{T_5 N_f}{4 \pi^2}\left( -\int d^{10}x \sin \theta \sin \tilde{\theta} e^{Y_9/2} \sqrt{-g_{(6)}} + \int (\sin \theta \sin\tilde \theta d \theta \wedge d \varphi \wedge d \tilde{\theta} \wedge d \tilde{\varphi}) \wedge C_{(6)}\right).
\end{equation}
Plugging our ans\"atze into the action and integrating over all coordinates except $\rho$ and dropping the overall volume factor we a get a one-dimensional action with Lagrangian
\begin{equation}\label{eq:lagrangian}
L= \sum_{i,j} G_{ij}(Y) Y_i'Y_j'-U(Y)\equiv T-U.
\end{equation}
We follow \cite{Gubser:2001eg} and express the $Y_i$ in terms of nine other functions to make $G_{ij}$ diagonal
\begin{equation}\label{eq:Ys}
\begin{array}{ccc}
  Y_1=e^{2z-6x}, & Y_2=e^{2z+2x}, & Y_3=e^{10y-2z+2l}, \\
  Y_4=e^{2y-2z+2p+2q}, & Y_5=e^{2y-2z+2p-2q}, & Y_6=e^{2y-2z-8p}, \\
  Y_7=a, & Y_8= b, & Y_9=\phi.
\end{array}
\end{equation}
One finds that
\begin{eqnarray}
T&=&e^{-l}\left(5y'^2-3x'^2-2z'^2-5p'^2-q'^2 -\frac14\,e^{-4q} a'^2-\frac{N_c^2}{64} \, e^{\phi+4z-4y-4p} b'^2-\frac18 \phi'^2 \right)\, , \nonumber \\
U&=&\frac18\,e^{l} \bigg[e^{8y}\left\{e^{-12p}\, [e^{4q}+e^{-4q}(a^2-1)^2+2a^2(1- e^{10p-2q})^2] -8e^{-2p}\cosh\ 2q \right\}  \nonumber \\
&+& \! \! \frac{N_c^2}{16} e^{\phi+4z+4y+4p} \! \left\{e^{4q} \! + e^{-4q} \! \left(\!a^2 \! \! - \! 2a b + \! 1 \! -\frac{N_f}{N_c} \! \right)^2 \! \!
\! + 2(a \! - \! b)^2 \! \right\} \! + \! N_c e^{\phi/2 + 2z + 6y - 4p} \bigg].
\end{eqnarray}
Note that the $N_c$ dependence can be absorbed by shifting the dilaton, which we will do from now on. Since $l$ has no kinetic term, it is a pure gauge degree of freedom reflecting the remaining reparametrization invariance. Varying with respect to $l$ one can set it to any value. To make contact with the zero temperature solutions in \cite{Caceres:2007mu}, we want to have $b=0$ which implies $a=-\sqrt{1-e^{4q}}$.\\
The equation of motion for $x$ can be integrated once to get
\begin{equation}
x'= - \frac14 e^{8 f_0} d \, e^l,
\end{equation}
where the constant prefactor is chosen to simplify later expressions.\\
By comparing the equations of motion for $z$ and $\phi$ we find
\begin{equation}
\phi' = 4z' + c_\phi e^l = 4 z' + c_x x',
\end{equation}
where the constant $c_x=-\frac{4 c_\phi}{ e^{8 f_0} d}$. We can integrate this to get
\begin{equation}\label{eq:dilaton}
\phi= 4 z + c_x x,
\end{equation}
where the integration constant can be set to zero by shifting the dilaton $\phi$. To make further contact with the solutions discussed in
\cite{Caceres:2007mu} we set $\frac{N_f}{N_c}=2$ and $l=-4(y+p-x)+\log{2}$. This leads to $Y_6=\frac{Y_1 Y_3}{4 Y_2}$. In the next section we solve the remaining equations to find analytic black hole solutions.

\subsection{Black hole solutions}\label{BHT}
In this section we present analytical black hole solutions that asymptote to the decoupled one-parameter family described in \cite{Caceres:2007mu}. To obtain these solutions we solve the remaining equations obtained from the Lagrangian \eqref{eq:lagrangian} for large $\rho$. Note that from \eqref{eq:metric} and \eqref{eq:Ys} we have that $Y_1/Y_2 = e^{-8x}$. So in order to find black holes we need to solve for the deformation parameter $e^{8x}$ to all orders in the radial variable since we need $e^{8x} \rightarrow \infty$ at the horizon.\\
In order to obtain solutions that are regular at the horizon, we have to choose the constant in \eqref{eq:dilaton} such that $c_x=4$. To compare with the results of \cite{Caceres:2007mu} we define
\begin{align}
Y_1=& e^{2f}e^{-8x}, \nonumber\\
Y_2=& e^{2f}, \nonumber\\
Y_3=& e^{2f} e^{2k} e^{8x},\nonumber\\
Y_4=& e^{2f} e^{2h},\nonumber\\
Y_5=& \frac{e^{2f} e^{2g}}{4},\nonumber\\
Y_6=& \frac{e^{2f} e^{2k}}{4},
\end{align}
where up to terms decaying more rapidly than $e^{-4\rho}$ we find
\begin{align}\label{eq:analytic}
e^{h-g} &=\frac12 - e^{-4\rho}, \nonumber \\
e^{h+g} &=1 + c e^{-\left(1+\sqrt{5}\right)\rho}+ 2 e^{-4\rho}, \nonumber \\
e^{2k} &=1 - \left(1+\sqrt{5}\right) c e^{-\left(1+\sqrt{5}\right)\rho} - 8 e^{-4\rho}, \nonumber \\
e^{2f} &=e^{2 f_0+\frac{\rho}{2}} \left(1 + \frac18 \left( -1+\sqrt{5} \right) c e^{-\left(1+\sqrt{5}\right)\rho} + \frac14 e^{-4\rho} \right), \nonumber \\
e^{8x} &= \frac{1}{1- 8 d e^{-2\rho}}.
\end{align}
Here $c \, \epsilon \, \mathbb{R}$ is a  free parameter distinguishing different solutions of the one-parameter family of decoupled solutions and $d$ corresponds to the temperature. For $d=0$ this reduces to a large $\rho$ expansion of the zero temperature solution found in \cite{Caceres:2007mu}. For $d<0$ we see that $e^{8x}$ remains finite and there is no black hole. However, for $d>0$ we have a horizon at $\rho_H =\frac{\log{8 d}}{2}$. Thus, for $d>0$, \eqref{eq:analytic} is a family of black hole solutions representing  the non-extremal deformation of the decoupled one parameter family of solutions presented in \cite{Caceres:2007mu}.

\subsubsection{Restrictions on parameter space}
\begin{figure}[!h]
\begin{center}
\includegraphics[width=5.3in]{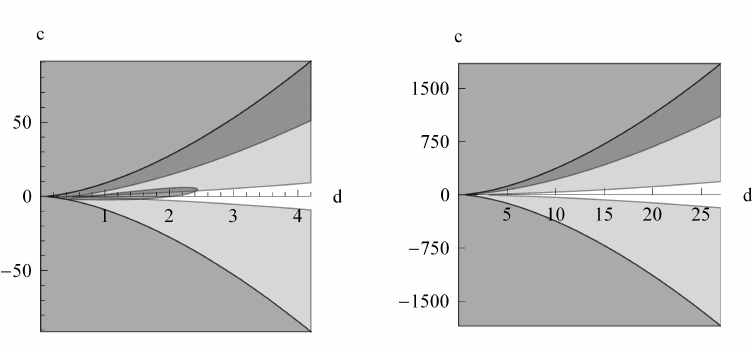}
\caption{ These plots show the $(d,c)$ plane. Our analytic black hole solutions are valid in the white region. The light gray region is excluded if we demand that $\left| (1+\sqrt{5}) c (8d)^{-\frac{1+\sqrt{5}}{2}} \right| < .1$. The Ricci scalar is larger than 1 in the dark gray areas. The plots also show the curve $|c|= \frac{(8d)^{\frac{1+\sqrt{5}}{2}}}{1+\sqrt{5}}$.}\label{fig:dcplane}
\end{center}
\end{figure}

To trust the analytic expression \eqref{eq:analytic} we need to satisfy two conditions on $c$ and $d$. Since we neglected terms decaying faster than $e^{-4\rho}$ and want the analytic expression \eqref{eq:analytic} to be valid all the way from the horizon $\rho_H$ to $\rho=\infty$ we need $\rho_H \gtrsim \frac14$ i.e. $d \gtrsim \frac{\sqrt{e}}{8}$. Furthermore, we have to demand that $c$ is not too large so
that the expansion for large $\rho$ is still valid. The most stringent constraint here comes from $e^{2k}$. We have to demand that $\left|
\left(1+\sqrt{5}\right) c e^{-\left(1+\sqrt{5}\right)\rho_H} \right| \ll 1$ i.e., $|c| \ll \frac{(8d)^{\frac{1+\sqrt{5}}{2}}}{1+\sqrt{5}}$. There is another constraint on the possible values of $c$ and $d$ allowed. In \cite{Casero:2006pt} it was shown that the curvature blows up, if we approach the singularity where the flavor branes end. The value of the radial coordinate at which the Ricci scalar becomes large depends on $c$ and $d$. Calculating the Ricci scalar for our analytic solutions we find additional regions that are excluded because the curvature becomes large and the supergravity approximation breaks down. In particular the region with $d\lesssim 2$ is excluded. Figure \ref{fig:dcplane}
shows the allowed regions in the $(d,c)$ plane.

\subsubsection{Temperature}
As long as these constraints on $c$ and $d$ are satisfied our analytic black hole solutions are valid. Calculating the black hole temperature observed at infinity we find
\begin{align}
T &= \frac{1}{4 \pi} \left. \sqrt{\frac{Y_1}{(\rho-\rho_H)^2 Y_3}} \right|_{\rho=\rho_H} \left. \sqrt{\frac{Y_3}{Y_1}} \right|_{\rho=\infty} = \left. \frac{1}{2 \pi \, e^{k}} \right|_{\rho=\rho_H} \nonumber\\
&= \frac{1}{2 \pi \sqrt{1 - \frac{\left(1+\sqrt{5}\right) c}{(8d)^{\frac{\left(1+\sqrt{5}\right)}{2}}} - \frac{1}{8 d^2}}},
\end{align}
where we have used the fact that for the decoupled solutions $\left. \frac{Y_3}{Y_1} \right|_{\rho=\infty}=1$. From this expression we see that for $c\geq0$ the temperature is always bigger than $T\geq \frac{1}{2\pi}$, which is obtained in the $d \rightarrow \infty$ limit. For $c<0$ we can do slightly better. If we set $\e =- (1+\sqrt{5}) c (8d)^{-\frac{1+\sqrt{5}}{2}}$, we find $T=\frac{1}{2 \pi \sqrt{1 +\e - \frac{1}{8 d^2}}}$. So for large $d$ we have $T \approx \frac{1}{2 \pi \sqrt{1 +\e}}$ and we recall that we can trust our analytic solution only for $\e \ll 1$. This should be compared with the result of \cite{Gubser:2001eg} where the minimal temperature possible was the Hagedorn temperature of the little string theory $T=\frac{1}{4 \pi}$. So we see that we have a similar situation and that at least for the analytic solutions the backreaction of the flavor branes has increased the minimal temperature we can have.

\subsubsection{Beyond analytic solutions}
\begin{figure}[!h]
\begin{center}
\includegraphics[width=6.5in]{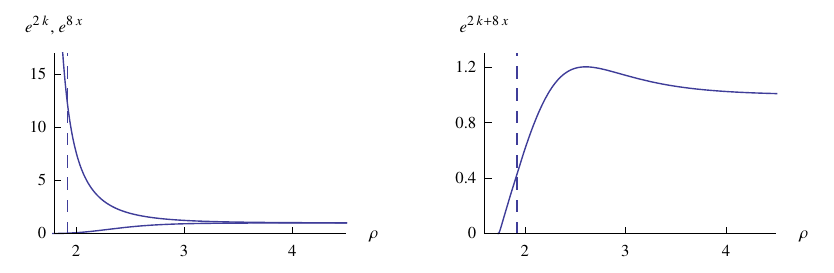}
\caption{This plot shows numerical results for $e^{2k}$ and $e^{8x}$ for $d=10$ and $c = \frac{(8d)^{\frac{1+\sqrt{5}}{2}}}{1+\sqrt{5}}$. One sees on the left that coming from large $\rho$ $e^{2k}$ starts going towards zero at $\rho \approx 3$ while $e^{8x}$ starts to blow up. The right graph shows the product $e^{2k+8x}$ which goes to zero. The dashed line at $\rho_c \approx 1.92$ indicates where the Ricci scalar becomes large.}\label{fig:c>0}
\end{center}
\end{figure}

We will now argue that there are no further black holes outside the regime of validity of our analytic solutions. First of all there are no black holes with a horizon at a small $\rho$ value. As we saw above and as was already noted in \cite{Caceres:2007mu} the curvature becomes large for $\rho \gtrsim 1$ which is still within thse validity of our analytic solutions. We have numerically examined the other two regions with $|c| \gtrsim
\frac{(8d)^{\frac{1+\sqrt{5}}{2}}}{1+\sqrt{5}}$ and the results agree with the expectations from the analytic solutions. One would generically expect that the solutions in these two regimes resemble the zero temperature results since the deformation parameter $d$ is relatively small compared to $c$. This is exactly what we find from the numerical solutions. \\
For $c \gtrsim \frac{(8d)^{\frac{1+\sqrt{5}}{2}}}{1+\sqrt{5}}$ one finds that $e^{2k}$ goes to zero and the curvature becomes large before we reach the horizon i.e., before $e^{8x}$ goes to infinity. This behavior could have been guessed from the expression for $e^{2k}$ in \eqref{eq:analytic} and the figure \ref{fig:dcplane}. A representative plot of this behavior is shown in figure \ref{fig:c>0}.

\begin{figure}[!h]
\begin{center}
\includegraphics[width=6.5in]{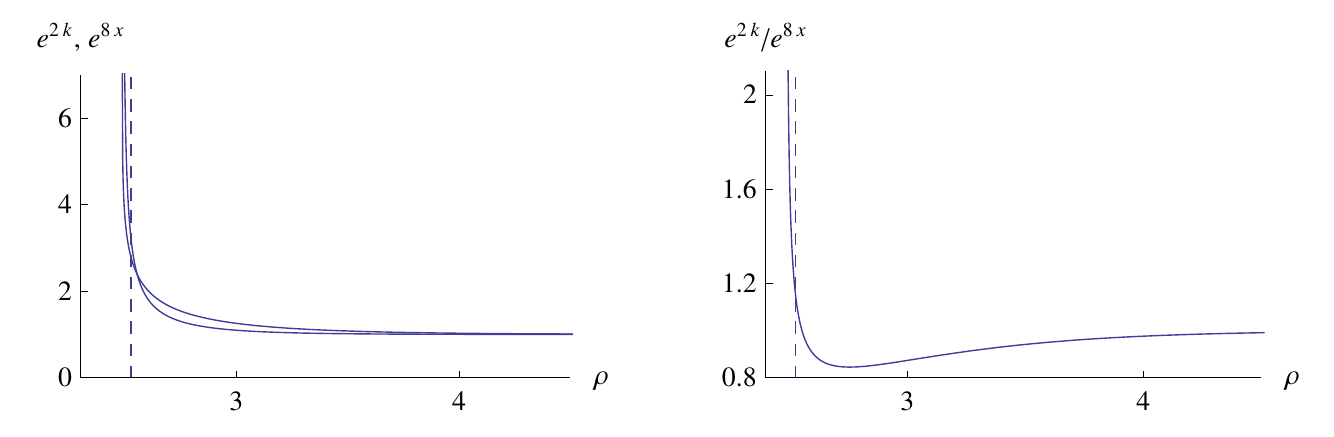}
\caption{This plot shows the numerical results for $e^{2k}$ and $e^{8x}$ for $d=10$ and $c = -\frac{(8d)^{\frac{1+\sqrt{5}}{2}}}{1+\sqrt{5}}$. Both $e^{2k}$ and $e^{8x}$ become large in this regime as can be seen in the left graph. The right plot shows $e^{2k}/e^{8x}$ and one sees that $e^{8x}$ is subdominant to $e^{2k}$ for small $\rho$. The dashed line at $\rho_c \approx 2.53$ indicates where the Ricci scalar becomes large.}\label{fig:c<0}
\end{center}
\end{figure}

\noindent
One sees that $e^{2k}$ becomes small and $e^{8x}$ starts to blow up for small $\rho$. The Ricci scalar also starts to blow up at $\rho_c \approx 1.92$ \footnote{The numerics depend to a certain extent on the exact value $\rho_\infty$ from which we start to integrate backwards. This does not affect the conclusion we reach.}. One finds that $e^{2k(\rho_c)} \approx 0.0356$ and $e^{8x(\rho_c)} \approx 12.1$ and at $\rho_f \approx 1.74$ we have $e^{2k(\rho_f)} \approx 2.60 \times 10^{-15}$ and $e^{8x(\rho_f)} \approx 134$. So as exemplified in the right graph in figure \ref{fig:c>0} we find that for $c \gtrsim \frac{(8d)^{\frac{1+\sqrt{5}}{2}}}{1+\sqrt{5}}$ the deformation $e^{8x}$ is subdominant and we do not have black holes in this regime.\\
For $c \lesssim -\frac{(8d)^{\frac{1+\sqrt{5}}{2}}}{1+\sqrt{5}}$ we find that $e^{2k}$ becomes very large and $e^{h+g}$ small. In figure \ref{fig:c<0} we plot the behavior of $e^{2k}$ and $e^{8x}$. We again find that the deformation $e^{8x}$ is subdominant.

\noindent
The behavior of $e^{2k}/e^{8x}$ in this regime is independent of the value of $d$ but it is interesting to study it as a function of $\e =- (1+\sqrt{5}) c (8d)^{-\frac{1+\sqrt{5}}{2}}$. One finds that $e^{8x}$ becomes dominant around $\e \approx .2$ which might still be within the validity of our analytic solution which required $\e \ll 1$.\\
So to summarize we find that outside the validity of our analytic solutions we have naked singularities. One might have expected this since the zero temperature solutions had those singularities and the deformation is small whenever the analytic solutions are not valid.

\section{Conclusions}
In previous work \cite{Caceres:2007mu} we presented new supergravity backgrounds incorporating the backreaction of $N_f = 2 N_c$ flavor D5 branes in
the CVMN background. In this note we find non-extremal generalizations of the solutions found in \cite{Caceres:2007mu}. For a particular range of parameters we find a one parameter family of analytic solutions with a regular horizon. Outside the range of validity of these analytic solutions we show, numerically, that the  solutions exhibit naked singularities. \\
The temperature of the regular solutions is always larger than the Hagedorn temperature of the LST. This is reminiscent of \cite{Gubser:2001eg} where the non-extremal CVMN background -without flavor branes- was studied and a similar behavior was found. Thus, the solutions found here seem to be dual to a LST above its Hagedorn temperature rather than to a finite temperature $\mathcal{N}=1$ $SU(N_c)$ gauge theory with a large number of fundamental matter.\\
We find that including flavor branes and their backreaction increases the minimal temperature from $T=\frac{1}{4\pi}$ \cite{Gubser:2001eg}, in the CVMN background without flavor, to $T = \frac{1}{2 \pi \sqrt{1+ \e}}, \, \e\ll1$ in the CVMN background with $N_f= 2 N_c$. Although our analysis is restricted to the special case $N_f = 2 N_c$ we believe that a similar conclusion can be reached for the more generic cases discussed in \cite{Casero:2006pt}, \cite{Casero:2007jj}, \cite{HoyosBadajoz:2008fw}.\\
We believe that the one parameter family of Hagedorn systems presented here can contribute to our understanding of Little String Theory. In particular, a detailed study of the thermodynamics of these solutions and the possible existence of a tachyonic mode that can be identified with the thermodynamic instability (\cite{Gubser:2000ec},\cite{Buchel:2005nt}) are issues we think deserve further study.

\section{Acknowledgments}
We  thank Matthias Ihl for collaboration at an initial stage of this work. E.C. thanks the Theory Group at the University of Texas at Austin, the Kavli Institute for Theoretical Physics and the Institute for Nuclear Physics at the University of Washington   for hospitality as well as the Department of Energy for partial support during the completion of this work. E.C. and R.F. thank the  Aspen Center for Physics for hospitality and partial support. The research of the authors is based upon work supported by the National Science Foundation under Grants No. PHY-0455649 and  PHY05-51164. The research of E.C. is also funded by CONACYT grant 50760. The work of T.W. is also supported by the German Research Foundation (DFG) within the Emmy-Noether Programme (Grant number ZA 279/1-2).

\end{document}